\documentclass[final,5p,times,twocolumn]{elsarticle}
\usepackage{amssymb}
\usepackage{lineno}
\usepackage[utf8]{inputenc}
\usepackage[T1]{fontenc}
\usepackage{epsfig}
\usepackage{amsmath}
\usepackage{amsfonts}
\usepackage{amssymb}
\usepackage{color}
\usepackage{graphicx}
\usepackage{url,hyperref}
\usepackage{latexsym}
\usepackage{longtable}
\usepackage{float}
\usepackage{epstopdf}
\usepackage{url}
\usepackage{caption}
\journal{Physics Letters B}
\begin{document}
%-------------------------------------------------------------------------------------------------
\begin{frontmatter}
%-------------------------------------------------------------------------------------------------
\title{Chaos in charged AdS black hole  extended phase space}
\author{M. Chabab $^{a}$}
\author {  \corref {mycorrespondingauthor}}
\cortext [mycorrespondingauthor] {Corresponding author}
\ead{mchabab@uca.ac.ma} 
\author{H. El Moumni $^{a,b}$}
%\ead{hasan.elmoumni@edu.uca.ma}
\author{S. Iraoui $^{a}$}
%\ead{s.iraoui@edu.uca.ma}
\author{K. Masmar $^{a}$}
%\ead{karima.masmar@edu.uca.ma}
\author{S. Zhizeh $^{a}$}
%\ead{sara.zhizeh@edu.uca.ac.ma}
%
\address{
$^{a}$ {High Energy and Astrophysics Laboratory, Physics Department, FSSM, Cadi Ayyad University, P.O.B. 2390 Marrakech, Morocco.}\\
$^{b}$ {LMTI, Department of Physics, Faculty of Sciences, Ibn Zohr University, Agadir, Morocco.}
}

%-------------------------------------------------------------------------------------------------
\begin{abstract}
We present an analytical study of chaos in a charged black hole in the extended phase space in the context of  the Poincare - Melnikov theory.  Along with some background on dynamical systems, we compute the relevant Melnikov function and find its zeros. Then we analyse these zeros either to identify the temporal chaos in the spinodal region, or to observe spatial chaos in the  small/large black hole equilibrium configuration. As a byproduct, we derive a constraint on the Black hole' charge required to produce chaotic behaviour.  To the best of our knowledge, this is the first endeavour to understand the correlation between chaos and phase picture in black holes. 
 
\end{abstract}
%-------------------------------------------------------------------------------------------------

%-------------------------------------------------------------------------------------------------
\begin{keyword}
Chaos,  AdS black holes,  Phase transitions.
\end{keyword}
%-------------------------------------------------------------------------------------------------
\end{frontmatter}
%

%-------------------------------------------------------------------------------------------------

\section{ Introduction.} 
One of the crowning achievements of the Golden Age of General Relativity was the discovery that black holes exhibit thermodynamical properties, providing a comprehensive treatment of the study of their phase transitions \cite{hawking}. Recently, one treats the cosmological constant as a thermodynamic variable (pressure) in a so-called extended phase space the whole black hole system has been mapped to the van der Waals fluid system successfully in many avenues with the fact that the $P-V$ diagram of the black holes at constant temperature and charges \cite{KM,our} consolidating further  the analogy between small/large black hole with the Van der Waals liquid/gas phase transitions \cite{our3,our4,our5,our6,our7,holo,our8,moiplb}.These developments have once again put black holes at the center of theoretical physics.  

The chaotic behaviour is ubiquitous in Nature,  black hole physics is not an exception \cite{Bombellitf, Manuele, Letelier}.  In these research,  the chaoticity is generally associated with inhomogeneous cosmological models \cite{Monerat:1998tw, Felder:2006cc}, with the aim to probe the chaotic behaviour of geodesic motion in black hole background \cite{Dubeibe:2007xka, Gair:2007kr, Chen:2016tmr}.
Although a vast literature on chaotic phenomenon around black holes exists, to our knowledge there are no studies that triggered a debate on the possible correlation chaos has to the thermodynamics of a black hole in an extended phase space.  The key idea of this letter is to elucidate this issue using the Melnikov method, a powerful analytical approach to locate chaotic features \cite{Melnikov-Holmes} to disclose the deep connection between charged AdS black holes and Van der Waals fluid system from a microscopic structure point of view.    
  
Following this line of inquiry, and  after a brief overview of the Melnikov Method, we present the main thermodynamic features of a charged AdS black hole.  Then we study the temporal chaos and show when it occurs in the spinodal region. Lastly, we re-compute the Melnikov function for identifying spatial chaotic behavior and prove that the homoclinic orbits become chaotic in the equilibrium configurations like those observed in Van der Waals fluid \cite{Slemrod,Felderhof, Widom}. Our hope is that these results can eventually help to figure out how to reach an understanding of the chaos with respect to the phase picture of black hole in an extended phase space.   

%%%%%%%%%%%%%%%%%%%%%%%%%%%%%%%%%%%%%%%%%%%%%%%%%%%%%%%
{\em The Melnikov method.} The Melnikov method  is an intrigued tool to perform analytical studies of chaotic systems\cite{Holmes1979,Holmes1990}. It is  based on the evaluation of the so called Melnikov integral, along  the  unperturbed  homoclinic orbit or the Melnikov function, the simple  zero points of this function are related to a discrete dynamical system \cite{Smale} which generates a certain class of chaos. More precisely, for a given  dynamical system,
   \begin{equation}\label{dynamic1}
  \dot{x}=f(x)+\epsilon g(x,t), \quad  x \in \Re^{2n} \; \text{and} \;
 \epsilon \ll 1, \end{equation}
 we assume that for  $\epsilon=0$ the system has a homoclinic orbit $\gamma_0(t)$  to the hyperbolic fixed point  $(0,0)$.  The key idea behind Melnikov method is to measure the distance between the invariant manifolds along the homoclinic orbit of the unperturbed vector field  with the assumptions that the perturbation is time  periodic and small. This distance is calculated using the Melnikov function  defined by,
 \begin{equation}
 M(t_{0})=\int_{-\infty}^{+\infty} \text{f}^{T}\left( \gamma_{0}(t-t_{0})\right)  \textbf{J}_n \text{g}\left( \gamma_{0}(t-t_{0}),t\right)dt,
 \end{equation}\\
 with,
 \begin{equation}
 \textbf{J}_{n=2}=\begin{pmatrix} 0 & 1 & 0 & 0 \\-1 & 0 & 0 & 0\\0 & 0 & 0 & 1 \\ 0 & 0 & -1 & 0 	\end{pmatrix}
\quad \text{and} \quad
\textbf{J}_{n=1}=\begin{pmatrix} 0 & 1  \\-1 & 0
	\end{pmatrix},
 \end{equation}
where the subscript $1$ and $2$ represent the number of degrees of freedom considered for the temporal chaos and spatial chaos respectively.
\\
 In the subsequent sections, we will use the Melnikov method to prove the occurrence  of either temporal chaos in the spinodal region, or a spatial chaos in the Van der Waals-like black hole equilibrium configuration.
 
 %%%%%%%%%%%%%%%%%%%%%%%%%%%%%%%%%%%%%%%%%
\section{ Temporal chaos in the spinodal region.} 
Here, we  investigate the effect of a small periodic perturbation about a subcritical 
temperature when the charged-AdS black hole is initially quenched in the unstable spinodal region.

The first step towards  our investigation 
 is to consider a spherically symmetric four dimensional metric  element describing the charged AdS black hole solution \cite{KM}
\begin{equation}
ds^2=-f(r)dt^2+\frac{dr^2}{f(r)}+ r^2d\Omega_2^2,
\end{equation}
whith,
\begin{equation}
f(r)=1-\frac{2 M}{r}+\frac{Q^{2}}{r^{2}}+\frac{\Lambda r^{2}}{3}.
\end{equation}
The parameters  $ M $ and $ Q $ read as  the ADM mass and charge of the black hole respectively, while $d\Omega_2^2$ stands for  the line element of the unit 2-sphere. The extended phase space is based on 
 the proposal that thermodynamic pressure is identified with the cosmological constant $\Lambda$ as, 
\begin{equation}
p=-\frac{1}{8\pi}\Lambda.
\end{equation}
Hence, the thermodynamical first law of such black hole can be written as,
\begin{equation}
d M= TdS + \Phi dQ+ V dp,
\end{equation}
where $V$ and  $\Phi$ represent the thermodynamic volume and the electric potential respectively.

In this context, the Hawking temperature is given by,
\begin{equation}
T=\left.\frac{f'(r)}{4\pi}\right|_{r=r_{+}}=\frac{1}{4\pi r_{+}}(1-\frac{Q^{2}}{r_{+}^{2}}-r_{+}^{2}\Lambda).
\label{eqq3}
\end{equation}
We can also derive the equation of state \cite{KM,our},
\begin{equation}\label{state}
p(\upsilon,T)=\frac{T}{\upsilon}-\frac{1}{2\pi\upsilon^{2}}+\frac{2Q^{2}}{\pi\upsilon^{4}}.
\end{equation} 
The parameter $\upsilon$ denotes the specific volume which is related to the event horizon radius via the formula $\upsilon=2  r_{+}$.
Eq. $\ref{state}$ reveals the  existence 
of some $p-\upsilon$ critical behaviour, similar to the 
Van der Waals liquid-gas one. The phase transition, of second order, shows up at  the following critical coordinates,
\begin{equation}
T_c=\frac{\sqrt{6}}{18 \pi Q},\quad \upsilon_c=2\sqrt{Q}, \quad p_c=\frac{1}{96 \pi Q^2}.
\end{equation}

  \begin{figure}[h]
	\begin{center}
		\includegraphics[scale=1]{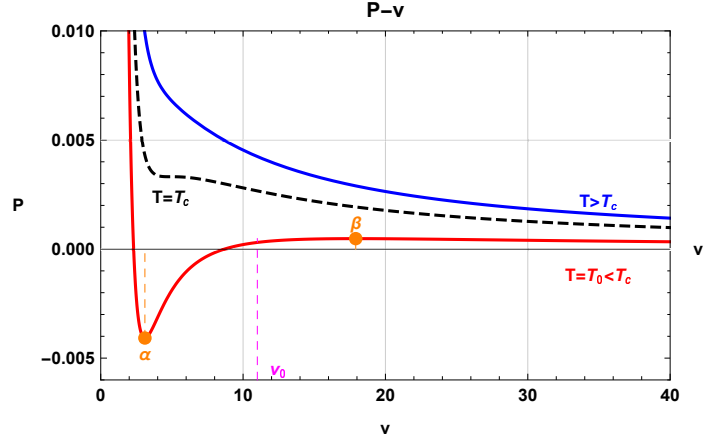}
		\caption{RN-AdS isotherms for $ Q=1 $.}
		\label{fig1}
	\end{center}
\end{figure}

For a temperature $T_0$ below the critical one,  we can  divide the specific volume  domain $\upsilon \in [0,\infty[$ into three regions, as illustrated in  Fig.\ref{fig1}:

 \begin{itemize}
	\item  $  [0,\alpha[ $: Corresponds to the  small black hole domain where $ \frac{\partial p(\upsilon,T_{0)}}{\partial \upsilon }<0$.
	\item $  [\alpha,\beta] $: The  spinodal region, where the small and large black hole phase coexist. In this region the factor   $ \frac{\partial p(\upsilon,T_{0)}}{\partial \upsilon }$ is positive. The two points $\alpha$ and $\beta$ are determined by $ \left. \frac{\partial p(\upsilon,T_{0)}}{\partial \upsilon }\right|_{\upsilon=\alpha}  =\left. \frac{\partial p(\upsilon,T_{0)}}{\partial \upsilon }\right|_{\upsilon=\beta} =0$.
	\item $  [\beta,\infty]$: The large black hole domain,  similarly to the small black hole region, refers  to a negative  factor $ \frac{\partial p(\upsilon,T_{0)}}{\partial \upsilon }$.
\end{itemize}
 
 It is worth noting that the point $\upsilon_{0}$ corresponds to  $ \frac{\partial^2 p}{\partial v^2}= 0$, on the graph of the $ T_{0} $ isotherm.

The charged AdS black hole flow is thought of as taking place along the x-axis in a tube of unit cross section of fixed volume. For the perturbation, we consider a specific volume $\upsilon_{0}$  in the instable spinodal region associated  to the isotherm $T_{0}$ and we will examine  the effect of small time-periodic  fluctuation of the  absolute temperature about $ T_{0} $, 
\begin{equation}
T=T_{0}+\epsilon \gamma cos(\omega t)cos(\tt M), 
\label{eq1}
\end{equation}
with $\epsilon \ll 1$. 

Our objective is to show, via Melnikov method, how the loss of stability of $\upsilon_{0} $ is accompanied by a temporal chaos.

We denote the Eulerian coordinate of a reference system by $x_{1}$. Then, the mass ${\tt M}$ of a column of black hole of  unit cross section between the reference  $x_{1}$ and the general Eulerian coordinates $ x $ parallel to r-direction is,
\begin{equation}
{\tt M}=\int_{x_{1}}^{x}\rho(\zeta,t)d\zeta,
\end{equation} 
where $\rho(x,t)$ is the density at position $x$ and time $t$. Notice also that $\rho(x(\tt M,t),t)^{-1}=x_{{\tt M}}({\tt M},t)=$$\upsilon$, where $ \upsilon $ denotes the specific volume.
 
Next, we study the thermodynamic phase transition exhibited in one dimensional thermal compressible RN-AdS black hole flow that can be described in Lagrangian coordinates by the following system:
\begin{equation}\label{eq11}
\left\{\begin{array}{c}\frac{\partial \upsilon}{\partial t}=\frac{\partial u}{\partial \tt M} \\ \\\frac{\partial u}{\partial t}=\frac{\partial \textbf{T}}{\partial \tt M}\end{array}\right.,
\end{equation}
where $u$ is the velocity and $\textbf{T}$  the stress tensor defined in Korteweig's theory \cite{Felderhof,Widom} as,
\begin{equation}
\textbf{T}=-p(\upsilon,T)+\mu \frac{\partial u}{\partial \tt M}-A \frac{\partial^2 \upsilon}{\partial \tt M^2},
\label{eq22}
\end{equation}\\
where $ p $ is the pressure given by Eq. \eqref{state}, and $ A $  a positive constant. The viscosity $\mu$ denotes an assumed positive constant.
By substituting \eqref{eq22} into \eqref{eq11}, we find that $x$ satisfies the equation:
\begin{equation}
\frac{\partial^2 x}{\partial t^2}=-\frac{\partial p(\upsilon,T)}{\partial \tt M}+\mu \frac{\partial^3 x}{\partial t\partial \tt M^2 } -A \frac{\partial^4 x}{\partial \tt M^4 }.
\label{eq5}
\end{equation}
Introducing the change of variables $ \bar{{\tt M}}=q {\tt M} $, $ \bar{t}=q t $, $ \bar{x}=q x $,  $ \mu=\epsilon \mu_{0} $, we can rewrite Eq. \eqref{eq5} 
as,
\begin{equation}
	\frac{\partial^2 \bar{x}}{\partial  \bar{t}^2}=-\frac{\partial p(\upsilon,T)}{\partial \bar{{\tt M}}}+\epsilon \mu_{0} q \frac{\partial^3 \bar{x}}{\partial \bar{t}\partial \bar{{\tt M}}^2 }-A q^{2} \frac{\partial^4 \bar{x}}{\partial \bar{{\tt M}}^4 }.
	\end{equation}
	For convenience, we omit the overbars in the subsequent part of this work.

In this case, the Hamiltonian is given by:
\begin{equation}
H=\frac{1}{\pi}\int_{0}^{2\pi}\left[\frac{u^{2}}{2}+\mathcal{F}(\upsilon,T)+\frac{A q^{2}}{2}\left( \frac{\partial \upsilon}{\partial \tt M}\right) ^{2}\right]d{\tt M},
\label{eq2}
\end{equation}
where $\mathcal{F}(\upsilon,T)$ is the free energy, 
\begin{eqnarray}
	\mathcal{F}(\upsilon,T)&=&-\int_{V_{0}}^{V} p(s,T) ds\\ \nonumber
	&=&\frac{\upsilon}{4}-\pi T \frac{\upsilon^{2}}{4}+\frac{ Q^{2}}{\upsilon}-\frac{\upsilon_{0}}{4}+\pi T \frac{\upsilon_{0}^{2}}{4}-\frac{ Q^{2}}{\upsilon_{0}},
	\end{eqnarray}
where $V=\frac{\pi\upsilon^{3}}{6}$ is the thermodynamic volume for the charged-AdS black hole.
At the equilibrium, with $ \upsilon = \upsilon_ {0} $ and $ u = 0 $, we first develop $\upsilon$ and $u$ in the Fourier series sinus and cosine, respectively. Then after performing a Taylor series expansion of $\frac{1}{\upsilon}$ about $\upsilon_0$ (up to $ \mathcal{O}(\frac{1}{\upsilon^{4}})) $, we can rewrite the Hamiltonian \eqref{eq2} as,
\begin{eqnarray}\nonumber
	H_{2}(x,u)&=&\frac{u_{1}^2}{2}+\frac{u_{2}^2}{2}+\frac{1}{2} A q^2 x_{1}^2-\frac{1}{4} \pi  T_{0} x_{1}^2+\frac{ Q^2
		x_{1}^2}{\upsilon_{0}^3}\\\nonumber
	& &+\frac{3 Q^2 x_{1}^4}{4 v_{0}^5}-\frac{3 Q^2 x_{1}^2 x_{2}}{2\upsilon_{0}^4}+2 A q^2
	x_{2}^2-\frac{1}{4} \pi  T_{0} x_{2}^2\\
	& &+\frac{ Q^2 x_{2}^2}{\upsilon_{0}^3}+\frac{3 Q^2 x_{1}^2
		x_{2}^2}{\upsilon_{0}^5}+\frac{3 Q^2 x_{2}^4}{4 \upsilon_{0}^5}\\
	& &-\frac{1}{2} \pi  \upsilon_{0} x_{1} \gamma  \epsilon  \cos (\omega t)-\frac{1}{4} \pi  x_{1} x_{2} \gamma  \epsilon  \cos ( \omega t), \nonumber
	\end{eqnarray}
where $(x_{1}, x_{2})$ and $(u_{1}, u_{2})$ represent the position and velocity of the two first modes respectively. Consequently we derive the following dissipative Hamiltonian structures: 
\begin{equation}
	\left\{
	\begin{aligned}
	\dot{x}_{1}=&\frac{\partial H_{2}}{\partial u_{1}}=u_{1}&\\
	\dot{x}_{2}=&\frac{\partial H_{2}}{\partial u_{2}}=u_{2}&\\
	\dot{u}_{1}=&-\frac{\partial H_{2}}{\partial x_{1}}-\epsilon\mu_{0}qu_{1}=-A q^2 x_{1}+\frac{\pi  T_{0} x_{1}}{2}-\frac{2 Q^2
		x_{1}}{\upsilon_{0}^3}&\\
	&-\frac{3 Q^2 x_{1}^3}{\upsilon_{0}^5}+\frac{3
		Q^2 x_{1} x_{2}}{\upsilon_{0}^4}-\frac{6 Q^2 x_{1}
		x_{2}^2}{\upsilon_{0}^5}&\\&+\frac{1}{2} \pi  \upsilon_{0} \gamma  \epsilon 
	\cos (\omega t)+
	\frac{1}{4} \pi  x_{2} \gamma  \epsilon  \cos (t w)-\epsilon\mu_{0}qu_{1}&\\
	\dot{u}_{2}=&-\frac{\partial H_{2}}{\partial x_{2}}-4\epsilon\mu_{0}qu_{2}=\frac{3 Q^2 x_{1}^2}{2\upsilon_{0}^4}-4 A q^2 x_{2}+\frac{\pi 
		T_{0} x_{2}}{2}&\\
	&-\frac{2 Q^2 x_{2}}{\upsilon_{0}^3}-\frac{6 Q^2
		x_{1}^2 x_{2}}{\upsilon_{0}^5}-\frac{3 Q^2
		x_{2}^3}{\upsilon_{0}^5}+\frac{1}{4} \pi  x_{1} \gamma  \epsilon 
	\cos (\omega t)&\\&-4\epsilon\mu_{0}qu_{2}&
	\end{aligned}
	\right.
	\label{eq7}
	\end{equation}
	
In the next step, we look for the homoclinic orbits. To this end we rewrite the system  \eqref{eq7} in the form of  \eqref{dynamic1} under the following form,
	\begin{equation}
	\dot{z}=f_{0}(z)+\epsilon f_{1}(z,t),\;\; with \quad z = (x_ {1}, x_ {2}, u_ {1}, u_ {2}) ^ {T}.
	\label{eq3}
	\end{equation}

The Jacobian linearization of the original nonlinear system \eqref{eq3} is given by:
\begin{equation}
	\textbf{J}=\begin{pmatrix} 0 & 0 &  1 & 0 \\0 & 0 & 0 & 1\\-A q^2+\kappa& 0 & -\epsilon\mu_{0}q & 0 \\ 0 & -4 A q^2+\kappa& 0 & -4\epsilon\mu_{0}q
	\end{pmatrix},
	\end{equation}
with $\kappa=\frac{\pi T_{0}}{2}-\frac{2 Q^2}{\upsilon_{0}^3}$. Its  eigenvalues are:
\begin{eqnarray}\nonumber
\lambda_{1,2}&=&\frac{-\epsilon\mu_{0}q}{2}\pm\frac{1}{2}[\epsilon^{2}\mu_{0}^{2}q^{2}-4(A q^2-\frac{\pi  T_{0}}{2}+\frac{4 Q^2}{\upsilon_{0}^3})]^{\frac{1}{2}},\\ \nonumber
\lambda_{3,4}&=&-2\epsilon\mu_{0}q\pm[4\epsilon^{2}\mu_{0}^{2}q^{2}-(4A q^2-\frac{\pi  T_{0}}{2}+\frac{4 Q^2}{\upsilon_{0}^3})]^{\frac{1}{2}}.
\end{eqnarray}
Under the following constraints, 
\begin{equation}
	\frac{\frac{\pi  T_{0}}{2}-\frac{2 Q^2}{\upsilon_{0}^3}}{4A} < q^{2}  < \frac{\frac{\pi  T_{0}}{2}-\frac{2 Q^2}{\upsilon_{0}^3}}{A},
	\end{equation}
 where $\epsilon$ is positive and sufficiently small, we can see that  the first mode is unstable when $ \lambda_{1}> 0 $ and $ \lambda_{2} <0 $, while  the higher modes are stable.
 
 Now, we focus on the  unperturbed system $\dot {z}(t) = f_{0}(z) $ which has a  two-dimensional invariant symplectic manifold containing the homoclinic orbit  and connecting the origin to itself \cite{Holmes1979},
\begin{equation}
		z_{0}(t)=\begin{pmatrix} \left( \frac{2a^{2}\upsilon_{0}^{5}}{3Q^{2}}\right) ^{\frac{1}{2}}sech(at)\\0\\a^{2}\left( \frac{2\upsilon_{0}^{5}}{3Q^{2}}\right) ^{\frac{1}{2}}sech(at)tanh(at)\\0
		\end{pmatrix},
	\end{equation}
where the parameter  $a$ is given by,
\begin{eqnarray}
a^{2}&=&-A q^2+\frac{\pi  T_{0}}{2}-\frac{2 Q^2}{\upsilon_{0}^3},
\end{eqnarray}
so, we can now define the Melnikov function as
\begin{equation}
M(t_{0})=\int_{-\infty}^{+\infty}f^{T}_{0}(z_{0}(t-t_{0})) \textbf{J}_{n=2} f_{1}(z_{0}(t-t_{0}),t)dt,
\end{equation}
where $f_{0}(z_{0}(t-t_{0}))$ and $ f_{1}(z_{0}(t-t_{0}),t) $ are given by,
\begin{equation} 	
f_{0}(z_{0}(t-t_{0}))=\begin{pmatrix} -\frac{a^2 \sqrt{\frac{2v_0^5}{Q^2}} \chi \xi}{\sqrt{3}}\\\frac{a^3 \sqrt{\frac{2v_0^5}{Q^2}} \chi}{\sqrt{3}}-\frac{ a^3 Q^2
	\left(\frac{2v_0^5}{Q^2}\right){}^{3/2}
	\chi^3}{\sqrt{3} v_0^5}\\0\\a^2 \chi^2
\end{pmatrix}
\end{equation}
and
\begin{equation}f_{1}(z_{0}(t-t_{0}),t)=\begin{pmatrix} 0\\\frac{1}{2} \pi  \upsilon_{0} \gamma  
\cos (\omega t)
+\mu_{0}qa^{2}\left( \frac{2\upsilon_{0}^{5}}{3Q^{2}}\right) ^{\frac{1}{2}}\chi \xi\\0\\\frac{1}{4} \pi \left( \frac{2a^{2}\upsilon_{0}^{5}}{3Q^{2}}\right) ^{\frac{1}{2}}\gamma   \cos (\omega t) \chi
\end{pmatrix},
\end{equation}
where $\chi=sech(a(t-t_{0}))$ and $\xi=tanh(a(t-t_{0}))$.

Our objective is to use the explicit function for homoclinic orbit $z_{0}(t)$ to evaluate the Melnikov function. After straightforward calculations we obtain,
\begin{eqnarray}
M(t_{0}&=-\int_{-\infty}^{+\infty} \left[\frac{a^2 \pi  \upsilon_{0} \sqrt{\frac{2\upsilon_{0}^5}{Q^2}} \gamma  \cos (
	\omega t) \chi \xi}{2
	\sqrt{3}}\right.\left. +\frac{2a^4 q \upsilon_{0}^5 \mu_{0} \chi^2 \xi^2}{3 Q^2}\right] dt.
\end{eqnarray}

To calculate this integral, we used the residues formalism and show that the resulting Melnikov function can be cast into a simplified form,
\begin{equation}
M(t_{0})= q\mu_{0} \textbf{I}+ w \gamma \textbf{J} \sin \left(\omega t_0\right),
\end{equation}
with,
\begin{equation}\textbf{I}= -\frac{2 a^3  v_0^5 }{9 Q^2},\; \text{and} \;\; 
\textbf{J}=\frac{\pi ^2 v_0^{7/2} \text{sech}\left(\frac{\pi  \omega }{2 a}\right)}{\sqrt{3}
	Q}.
\end{equation}	
  Note that the Melnikov  function $M (t_ {0})$ has a simple zero only when the following condition is satisfied, 
  	 \begin{equation}\label{condition}
  	\rvert\frac{q\mu_{0}\textbf{I}}{\omega\gamma\textbf{J}}\rvert\leqslant 1.
  	 \end{equation} 
 Therefore, we can deduce that sufficiently small perturbations with $ \epsilon> 0 $ can generate chaos behavior due to temporal thermal  fluctuations.
   
Besides, by using Eq. \eqref{condition}, we  derive an important constraint on the black hole charge required that must be fulfilled to observe the chaotic features. 
\begin{equation}
\frac{\frac{2\sqrt{3}}{9} a^3 \mu_0 q v_0^{3/2} \cosh \left(\frac{\pi 
   \omega }{2 a}\right)}{ \pi ^2 \gamma  \omega }\leq Q.
\end{equation}
  
  Having obtaining the  essential features originating from  
  the effect of a small periodic perturbation, where the temperature is below the critical one,  when the charged-AdS black hole is initially quenched in the unstable spinodal region,
we will investigate in the next section the effects of small spatially periodic thermal variations of a Van der Waals-like black hole in the equilibrium configuration.
  
  %%%%%%%%%%%%%%%%  %%%%%%%%%%%%%%%% 
  \section{ Spatial Chaos  in the equilibrium configuration.}  Here, we consider the effect of a small spatially periodic perturbation in the equilibrium configuration with an  absolute temperature ($T_{0}<T_{c}$) expressed in the following form \cite{Slemrod},
   \begin{equation}
   T=T_{0}+\epsilon cos(qx).
   \label{eq1d}
   \end{equation}
   According to the Korteweig theory, the stress tensor is given by
   \begin{equation}
   \tau=-p(\upsilon,T)-A\upsilon^{''},
   \end{equation}	
where the notation $'=\frac{d}{dx}$ has been used. It is worth noting that,  for a static equilibrium with no body forces, we get $\frac{d\tau}{dx}=0$, so $ \tau= const = -B $ , where B is the ambient pressure. 

\begin{figure}[h]
   	\begin{center}
   		\includegraphics[scale=1]{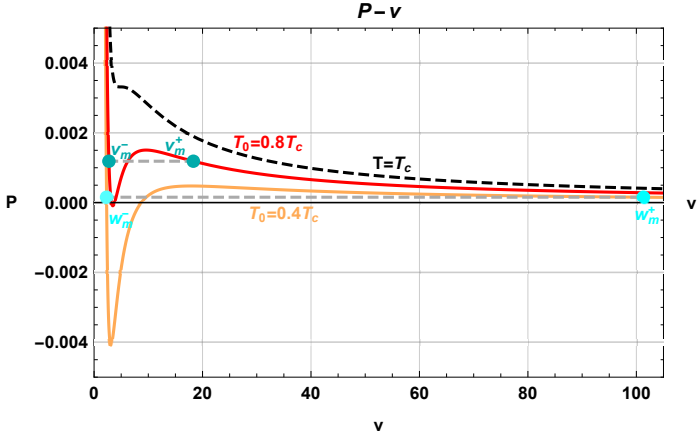}
   		\caption{RN-AdS isotherms with the  Maxwell equal area construction for $Q=1$  }\label{Fig2}
   	\end{center}
   \end{figure}

   Next step, we will focus on the homoclinic and heteroclinic orbits. For the unperturbed system ($ T=T_{0} $), we have:
  \begin{equation}
   \upsilon^{''}=B-p(\upsilon,T_{0}).
   \label{eq4}
   \end{equation}	
   The fixed points of the  non linear systems in Eq. \eqref{eq4} are the specific volumes corresponding to the pressure B in the diagram $ p-\upsilon $ shown in Fig. \ref{Fig2}.
   
  For illustration, we consider three isotherms with pressure $B$:  $ T = 0.8 T_{c}$ %, $ T = 0.6 T_{c}$ 
   and $T = 0.4 T_c$. For the corresponding specific volumes, we make use of the notation ($ v_{1} $, $ v_{2} $ and $ v_{3} $)   for  $ T = 0.8 T_{c}$,% ($ u_{1} $, $ u_{2} $ and $ u_{3} $) for  $T = 0.6 T_c$, and 
   while ($ w_{1} $, $ w_{2} $and $ w_{3} $) for $T = 0.4 T_c$. We also use the sign $+ $ and $-$ to denote the specific volume associated with Maxwell equal area construction,  
   \begin{equation}
   \int_{\upsilon_{m}^{-}}^{\upsilon_{m}^{+}}\left\lbrace {p(\upsilon,T_{0})-p(\upsilon_{m}^{-},T_{0})}\right\rbrace d\upsilon=0.
   \end{equation}
 
\begin{center}
\begin{figure}[!ht]
\begin{tabbing}
\hspace{9cm}\=\kill
\includegraphics[scale=1]{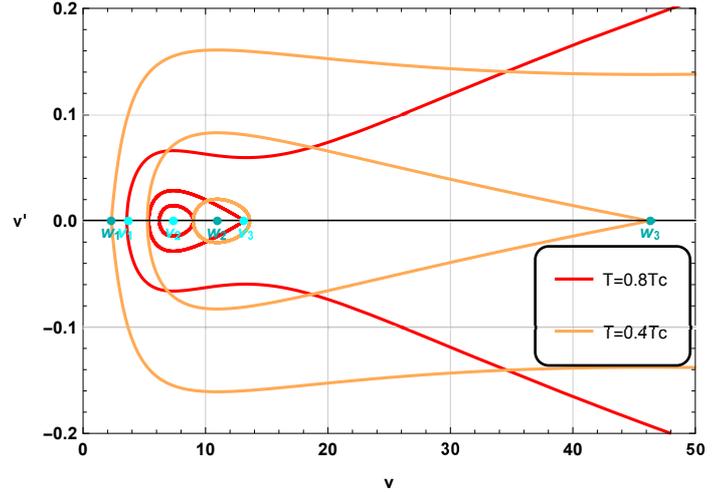}\\ \includegraphics[scale=1]{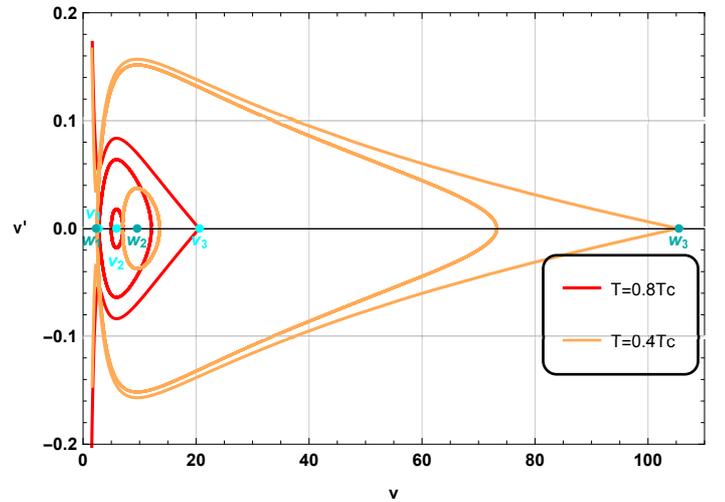} \\  \includegraphics[scale=1]{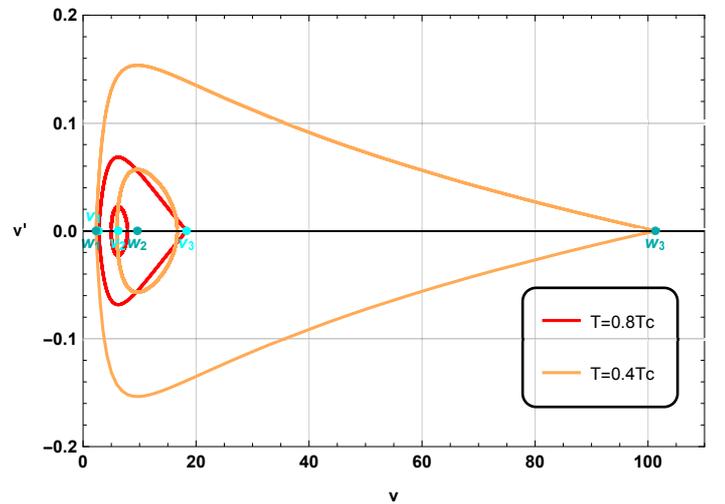}
\end{tabbing}
\vspace*{-.2cm} \caption{\footnotesize Phase structures for case 1 (Top), case 2 (Middle) and case 3 (Bottom).} \label{fig3}
\end{figure}
\end{center}

 Analysing  the differential equation \eqref{eq4} which describes one-degree-of-freedom Hamiltonian system,  we distinguish three different types of phase structure in the $v-v'$ phase plan as illustrated in Fig. \ref{fig3}. In all three cases, ($ v_{1} $, $ v_{3}$) or  ($ w_{1} $, $ w_{3} $)  are saddle points pairs, while $ v_{2} $ or  $ w_{2} $  are center ones, respectively.
  
  \begin{itemize} 
 \item{\bf Case 1}: The pressure $B$ is in the range $p(\upsilon^{-}_{m},T_{0}) <B<p(\beta,T_{0})$. 
 We can see that the  system has a homoclinic orbit connecting the stable and unstable manifolds of saddle point $ v_{3} $ ($ w_{3}$ ), as illustrated by the top panel of  Fig. \ref{fig3}.
   
 \item{\bf Case 2}:    $p(\alpha,T_{0}) <B<p(\upsilon^{+}_{m},T_{0}) $ represented by the middle panel of Fig. \ref{fig3},   the system has again a homoclinic orbit connecting the saddle point $ v_{1}$ ($ w_{1} $ ) to  itself, respectively.
   
\item{\bf Case 3}:     Here $B=p(\upsilon^{+}_{m},T_{0})=p(\upsilon^{-}_{m},T_{0})$: 
   For this case crresponding to the bottom panel of Fig. \ref{fig3}, we obtain instead  heteroclinic orbit connecting the states $ v_{1} $ to $ v_{3} $ and  $ w_{1} $ to  $ w_{3}$.
   \end{itemize}

   To summarise, the unperturbed system possess a homoclinic orbit for the first and second cases, whereas it has a heteroclinic orbit for case 3. These features allow to compute the Melnikov function for these orbits.
   
   Considering  the temperature expression given in Eq. \eqref {eq1d},  we can rewrite the  perturbed system as,
      \begin{equation}
   \upsilon^{''}=B-p(\upsilon,T_{0})-\frac{\epsilon cos(qx)}{\upsilon}.
   \label{eq6}
   \end{equation}
  We then express the Melnikov function as:
   \begin{equation}
   M(x_{0})=\int_{-\infty}^{+\infty}f(Z(x-x_{0})) \textbf{J}_{n=1} g(Z(x-x_{0}),x)dx.
   \end{equation}
   Setting $ \upsilon^{'}=w $,  the system \eqref{eq6} becomes:
   \begin{equation*}
   \left\{
   \begin{aligned}
   \upsilon^{'}&=w&\\
   w^{'}&=B-p(\upsilon,T_{0})-\frac{\epsilon cos(qx)}{\upsilon}&\\
   \end{aligned}
   \right.,
   \end{equation*}
   $ Z(x)=\left( \begin{array}{c}
   \upsilon_{0}(x) \\
   w_{0}(x) \\
   \end{array} \right) $ is the homoclinic orbit for the cases 1 and 2, and the heteroclinic orbit for the case 3. The expressions of the $f$ and $g$ functions are given by,
   \begin{eqnarray} 
    f(Z(x-x_{0}))&=&\left( \begin{array}{c}
   w_{0}(x-x_{0})\\
   B-p(\upsilon_{0}(x-x_{0}),T_{0}) \\\end{array} \right), \\
   g(Z(x-x_{0}),x)&=&\left( \begin{array}{c}
   0 \\
   -\frac{cos(qx)}{\upsilon_{0}(x-x_{0})} \\\end{array} \right). 
   \end{eqnarray}
   If we introduce the change of variable $X=x-x_0$, we can show that $ M(x_{0}) $,
   \begin{equation*}
   M(x_{0})=\int_{-\infty}^{+\infty}-\frac{w_{0}(x-x_{0})cos(q x)}{\upsilon_{0}(x-x_{0})}dx,
   \end{equation*}
 reduces to, 
   \begin{equation}
   M(x_{0})=-Lcos(qx_{0})+Nsin(qx_{0}).
   \end{equation}
   With:
   \begin{eqnarray}
   	 L&=&\int_{-\infty}^{\infty}\frac{w_{0}(X)cos(qX)}{\upsilon_{0}(X)}dX,\\
   	 N&=&\int_{-\infty}^{\infty}\frac{w_{0}(X)sin(qX)}{\upsilon_{0}(X)}dX.
   \end{eqnarray}
   Hence whatever the values of $L$ and $N$, the function $M (x_ {0})$ has simple zeros.  Consequently, for a sufficiently small $ \epsilon $ and a pressure $B$ in the range $p(\alpha,T_{0}) <B<p(\beta,T_{0}) $,  corresponding to the region where phase transition between small and large black hole occurs, an  equilibrium  configuration  will definitely  present a spatial chaos.\\
 
 %%%%%%%%%%%%%%%%%
\section{ Conclusion.} 
We have presented an analytical study of chaos of charged  Anti-de-Sitter black hole in extended phase space. More specifically, using the standard Poincare - Melnikov theory, we derived  the Melnikov function and its zeros, then we proved the existence of  temporal perturbation in the spinodal region. We also observed a spatial chaos in the small/large black hole equilibrium configuration. Furthermore, we have identified an upper limit on the black hole charge that prevents the occurrence of  the chaotic behaviour if exceeded.  Although a vast literature on chaotic phenomenon around black holes exists, to our knowledge this is the first time a new approach is provided to probe the phase picture from the point view of chaos dynamics.
 
 Lastly, needless to mention that we hope that this study can be extended to other backgrounds, including those with higher dimensions and higher derivative  gravity models.

\section*{References}

%\bibliography{mybibfile}

\end{document}